\documentclass[12pt]{iopart}

\usepackage{graphics}
\usepackage{color}

\begin{document}

\title[Graphene nanoring as a tunable source of polarized electrons]%
{Graphene nanoring as a tunable source of polarized electrons}

\author{J.\ Mun\'{a}rriz,$^1$ F.\ Dom\'{\i}nguez-Adame,$^1$  
P.\ A.\ Orellana$^2$ and A.\ V.\ Malyshev$^{1,3}$}

\address{$^1$ GISC, Departamento de F\'{\i}sica de Materiales, Universidad
Complutense, E-28040 Madrid, Spain}

\address{$^2$ Departamento de F\'{\i}sica, Universidad Cat\'{o}lica del Norte,
Casilla 1280, Antofagasta, Chile}

\address{$^3$ Ioffe Physical-Technical Institute, St-Petersburg, Russia}

\begin{abstract}

We propose a novel spin filter based on a graphene nanoring fabricated above a
ferromagnetic strip. The exchange interaction between the magnetic moments of
the ions in the ferromagnet and the electron spin splits the electronic states,
and gives rise to spin polarization of the conductance and the total electric
current. We demonstrate that both the current and its polarization can be
controlled by a side-gate voltage. This opens the possibility to use the
proposed device as a tunable source of polarized electrons.

\end{abstract}


\submitto{\NT}

\maketitle

\section{Introduction}

Graphene is a material with a combination of many remarkable properties, in
particular, large electron mobility and long spin-coherence lengths up to
several microns~\cite{Kane05,Tombros07,Yazyev08}. These features spurred the
interest in graphene as a material of choice for spintronic devices which
exploit both the charge and the spin degrees of freedom as the basis of their
operation. Geim and coworkers used soft magnetic NiFe electrodes to inject
polarized electrons into graphene and found spin valve effects~\cite{Hill06}.
Later, Cho \emph{et al.} performed four-probe spin-valve experiments on graphene
contacted by ferromagnetic Permalloy electrodes~\cite{Cho07}, observing a
switching of the sign in the four-probe nonlocal resistance, which indicates the
presence of a spin current between injector and detector. The drift of electron
spins under an applied dc electric field in  spin valves in a field-effect
transport geometry at room temperature was studied in Ref.~\cite{Josza08}. These
experiments were found to be in quantitative agreement with a drift-diffusion
model of spin transport. More recently, Dedkov \emph{et al.} proposed that the
Fe$_{3}$O$_{4}$ /graphene/Ni trilayer can also be used as a spin-filtering
device, where the half-metallic magnetite film was used as a detector of
spin-polarized electrons~\cite{Dedkov11}.

With the development of the nanoscale technology of graphene, a number of
nanodevices have been proposed to explore novel spin-dependent transport
phenomena. Spin filter effects in graphene nanoribbons with zig-zag edges were
investigated theoretically by Niu and Xing using a non-equilibrium Green
function method~\cite{Niu10}. They found a fully polarized spin current through
ferromagnetic graphene/normal graphene junctions, whose spin polarization could
be manipulated by adjusting the chemical potential of the leads. Ezawa
investigated similar effects in a system made of graphene nanodisks and leads,
where the magnetic moment of the nanodisk can be controlled by the spin
current~\cite{Ezawa09}. Guimar\~{a}es \emph{et al.} studied spin diffusion in
metallic graphene nanoribbons with a strip of magnetic atoms substituting carbon
ones in the honeycomb lattice~\cite{Guimaraes10}. They found that the system
behaves as a spin-pumping transistor without net charge current. More recently,
Zhai and Yang have shown that the combined effects of strained and ferromagnetic
graphene junctions can be used to fabricate a strain-tunable spin
filter~\cite{Zhai11}. All these findings open the possibility of designing
spintronic devices based on graphene nanostructures for memory storage and spin
diodes. For a more complete and detailed review on the electronic and spin
properties of mesoscopic graphene structures and control of the spin see, for
example, Ref.~\cite{Rozhkov11} and references therein.

Recently we have proposed a novel design of a quantum interference device based
on a graphene nanoring, in which all edges are of the same type to reduce
scattering at the bends~\cite{Munarriz11}. Electron transport in the device can
be controlled by a gate voltage applied across the nanoring between two side
electrodes. The relative phase of the electron wave function in the two arms of
the ring can be varied by the side-gate voltage, leading to constructive or
destructive interference at the drain, which results in conductance oscillations
and electric current modulation. In this paper we show that such controlled
interference can also be used to design an efficient spin filter device. The
latter can be achieved by depositing a ferromagnetic insulator below (or above)
the nanoring, in which case the combination of the exchange splitting due to the
interaction of the electron spin with the magnetic ions and the effect of the
side-gate voltage can result in a controllable spin-polarized electric current.

\section{Graphene nanoring: Model and formalism}

\begin{figure}[ht]
\centerline{\resizebox{0.6\columnwidth}{!}{\includegraphics{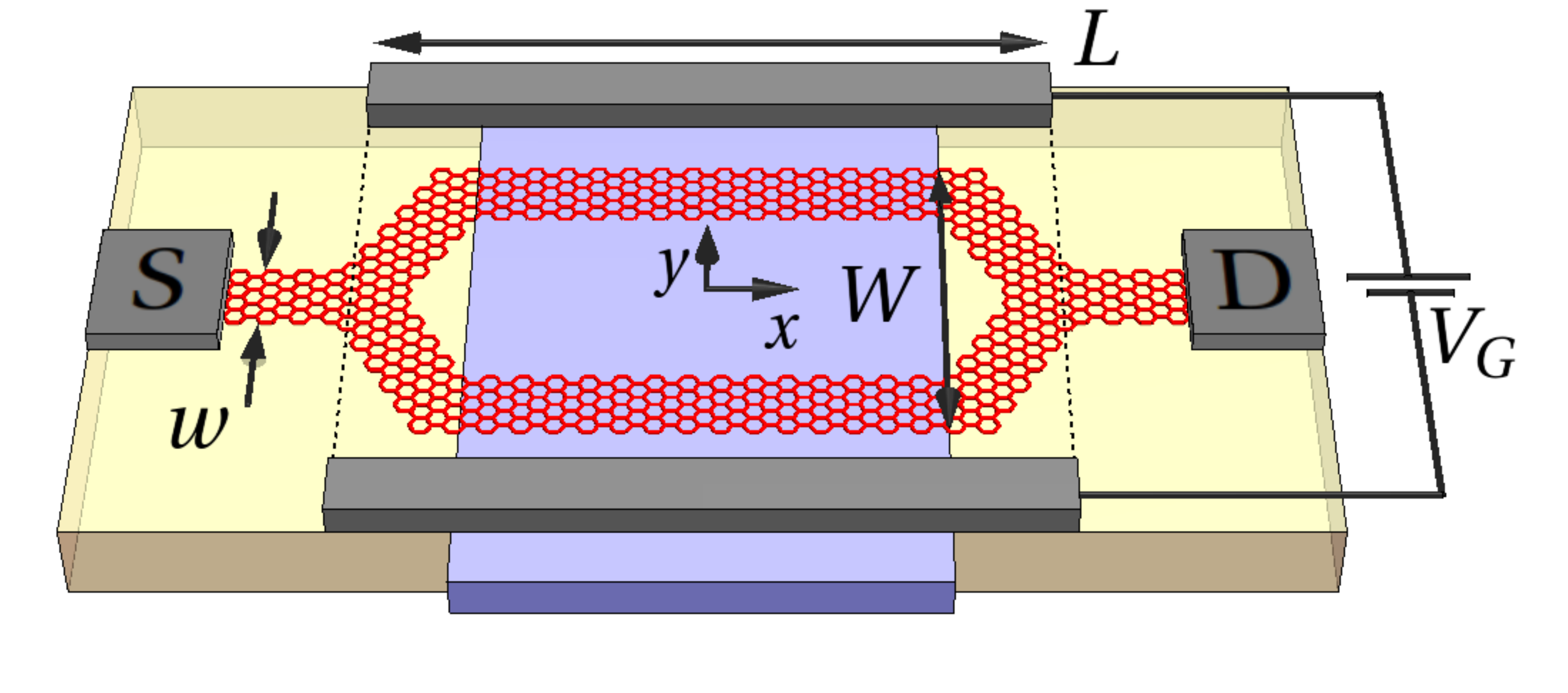}}}
\caption{
Schematic view of the graphene nanoring fabricated above a ferromagnetic
strip (shown as the blue bar in the figure). Source and drain terminals are
denoted as S and D respectively. Dimensions $w$, $W$ and $L$ are given in
the text.
}
\label{fig1}
\end{figure}

The quantum interference device introduced in Ref.~\cite{Munarriz11} consists of
a graphene nanoring with $60^\circ$ bends attached to two graphene nanoribbons
which, in turn, are connected to source and drain terminals, as shown in
figure~\ref{fig1}. The width of all nanoribbons is $w$. Two lateral electrodes
allow to apply a side-gate voltage.  The total length of the ring is $L$ while
the total width is $W$. Note that these dimensions should be large enough to
avoid dielectric breakdown at used source-drain and side-gate voltages. 

The dispersion relation of graphene nanoribbons is known to be very different
from that of the bulk graphene~\cite{Nakada96}. In particular, their band
structure has a width-dependent gap that turns out to be very sensitive to the
type of the nanoribbon edge (zig-zag or armchair)~\cite{Son06}.  In the most
general case, the edge type changes at a nanoribbon bend, which results in the
energy levels mismatch and can decrease the transmission and the current
considerably.  To reduce the back scattering at bends, we proposed to use
$60^\circ$ bends in order to keep all edges of the device of the same
type~\cite{Munarriz11}.  Recent studies indicate that atomic-scale precision
along the edges can be achieved experimentally~\cite{Jia09}, promising a
possibility to fabricate the proposed device.  Hereafter we restrict ourselves
to nanoribbons with armchair edges, for which the energy spectrum is centered
around $k=0$.  This type of nanoribbon was found to be the most advantageous one
for electronic transport because the transmission spectrum of such a system
presents wide bands of high transmission probability~\cite{Munarriz11}.

The applied side-gate voltage results in different energy shifts of the
electronic states in the two arms of the ring.  Thus, a charge carrier
injected from the source nanoribbon couples to different modes of the two
arms.  These two modes can interfere constructively or destructively at the
drain, giving rise to conductance and current modulation depending on the
side-gate voltage~\cite{Munarriz11}.  We note that this form of current
control relies on interference effects which depend quantitatively on
details of device geometry, material parameters, perturbations, such as
the disorder, etc.  However, the underlying principle of operation is very basic
and as long as there exist propagating modes in the two arms of the ring,
the control is expected to be feasible.  We will use therefore simple models
which grasp the main features of the different system components.

\begin{figure}[ht]
\centerline{\resizebox{0.8\columnwidth}{!}{\includegraphics{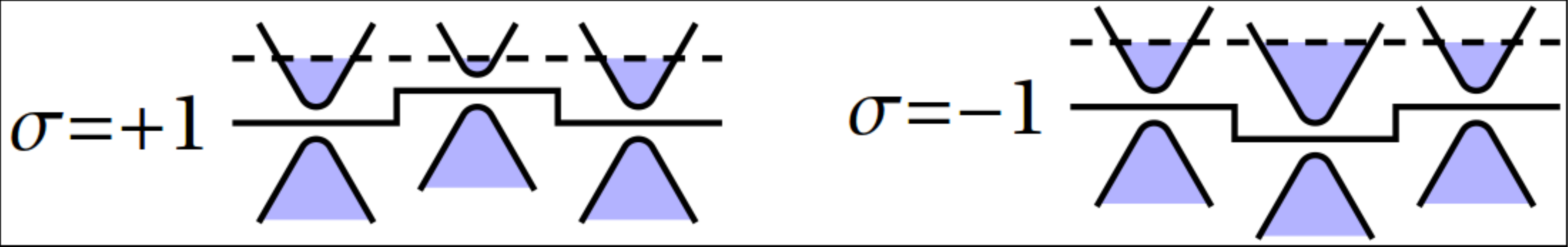}}}
\caption{
Schematic representation of the potential profile along one of the arms of
the ring for spin up (left) and spin down (right) states.  The potential
(solid line) is spin dependent in the middle section of the ring due to the
exchange splitting.  Dispersion relations of a nanoribbon with armchair
edges and band fillings up to the Fermi energy (dashed line) are also shown
schematically.
}
\label{figFermi}
\end{figure}

In the present design we consider a nanoring fabricated above a strip of a
ferromagnetic insulator, such as EuO. The exchange interaction between Eu$^{2+}$
ions and charge carriers can be described as an effective Zeeman splitting of
the spin sublevels~\cite{Haugen08}. 
%
%
This creates spin-dependent potential profiles along the arms
of the ring (see figure~\ref{figFermi}), so an injected electron couples to
different modes of the arms depending on its spin and therefore, for some
side-gate voltages, the interference governing the conductance can be
constructive for spin-up and destructive for spin-down states (or vice versa),
resulting in a spin polarized total current.

To model the device we have considered a simple tight-binding Hamiltonian of
a single electron in the $\pi$ orbitals of graphene within the
nearest-neighbor approximation
\begin{equation}
\mathcal{H}= \sum_{i}\epsilon_i |i \rangle\langle i| + 
\sum_{\langle i,j\rangle}V_{ij}|i\rangle\langle j|
+\sigma\,\Delta_\mathrm{ex}\sum_{i\in {\cal L}}|i \rangle\langle i| \ ,
\label{Hamiltonian}
\end{equation} 
where the site energy $\epsilon_i$ depends on the position of the $i$-th
carbon atom, in particular, due to the side-gate voltage.  Following
Ref.~\cite{Munarriz11} we use a simplified side-gate potential profile: if the
origin of the coordinate system is in the geometrical center of the ring (as
indicated in figure~\ref{fig1}), this potential is linear in the $y$ direction
for $|y|<W/2$ while in the $x$ direction it is (i) constant within the nanoring
area ($|x|\leq L/2$) and (ii) decays exponentially towards the two leads (for
$|x|\geq L/2$).  The contacts are assumed to be far enough from the ring, so
that the side-gate potential can be safely set to zero at the leads.  The total
potential drop between the outer edges of the two arms (separated by the
distance $W$) is denoted as $U_G$ and is referred to as the side-gate voltage
from now on.

The ferromagnet also affects the site energies $\epsilon_i$, shifting them
by the amount $\sigma \Delta_\mathrm{ex}$, where $\Delta_\mathrm{ex}$ is the
exchange splitting amplitude and $\sigma=+1$ ($\sigma=-1$) for spin up (spin
down) states.  Throughout the paper we use the value of
$\Delta_\mathrm{ex}=3\,$meV, which is of the order of the values known from
the literature~\cite{Haugen08,Zou09,Gu09}.  Because the characteristic
length scale of the exchange interaction is of the order of one monolayer
thickness,~\cite{Haugen08} we assume that only the sites which are in touch
with the ferromagnetic strip are affected by the interaction; the set of
such sites belonging to the longitudinal sections of the arms (see
figure~\ref{fig1}) is denoted as $\cal L$ in equation~(\ref{Hamiltonian}). 
Due to such local splitting, a spin up (down) electron propagating along one
of the arms is subjected to the potential with a rectangular barrier (well).
%
%
Such potential profile is shown schematically in figure~\ref{figFermi} for
the case of zero side-gate voltage.

The graphene lattice is known to undergo reconstruction at nanoribbon edges,
which affects the corresponding site energies $\epsilon_i$ and hoppings
$V_{ij}$~\cite{Son06}.  These effects are not expected to play a crucial
role for transport properties of realistic disordered samples (see
Ref.~\cite{Munarriz11} for details), and therefore we neglect them and
consider an undistorted honeycomb lattice with nearest neighbor coupling
$V_{ij}=-2.8\,$eV~\cite{Castro09}.

The quantum transmission boundary method~\cite{Lent90,Ting92} combined with the
effective transfer matrix approach~\cite{Schelter10} were used to calculate 
wave functions and spin-dependent transmission coefficients $T_\pm$ for spin
up~($+$) and spin down~($-$) electrons. These coefficients depend on the energy
of the injected carrier $E$ and the side-gate voltage $U_G$ and allow to 
obtain the degree of transmission polarization
\begin{equation}
P_T =\frac{T_{+}-T_{-}}{T_{+}+T_{-}}\ ,
\label{polarization}
\end{equation}
as well as spin polarized currents $I_\pm$. To calculate the latter we used
the Landauer-B\"{u}ttiker scattering formalism
\begin{equation}
I_\pm=\frac e h \int 
T_\pm(E,U_G)\,\Big[f(E,\mu_S)-f(E,\mu_D)\Big] dE\ ,
\label{eqIntensity}
\end{equation}
where
$
f(E,\mu)=\big\{\exp[(E-\mu)/k_B T)]+1\big\}^{-1}
$
is the Fermi-Dirac distribution, $k_B$ being the Boltzmann constant. Here we
assume that the source-drain voltage $V_\mathrm{SD}$ drops in the leads, which
agrees with recent experimental results~\cite{Venugopal10}, and model the effect
of this voltage drop as a shift of the Fermi level of the source, $\mu_S$, with
respect to that of the drain, $\mu_D$.  Then, one can calculate the total current
through the device
\begin{equation}
I ={I_{+}+I_{-}}\ ,
\label{I}
\end{equation}
and its polarization
\begin{equation}
P =\frac{I_{+}-I_{-}}{I_{+}+I_{-}}\ .
\label{Ipolarization}
\end{equation}
Finally, we note that equation~(\ref{eqIntensity}) is valid in the one-mode
approximation~\cite{Buttiker85}.  Hereafter, we assume that our system is
operating in the one-mode regime, which implies that the lateral quantization
energy (due to the finite nanoribbon width $w$) is much larger than the
source-drain voltage $V_\mathrm{SD}$ and the temperature $T$.  Further
requirements for the latter two parameters are discussed below.

\section{Results and discussion}

The sample that we consider is a nanoring made of nanoribbons of width
$w=15.5\,$nm with armchair edges. The full length of the ring is $L=179\,$nm
while the full width is $W=108\,$nm (see figure~\ref{fig1} for the schematics of
the device).

\begin{figure}[ht]
\centerline{\resizebox{0.6\columnwidth}{!}{\includegraphics{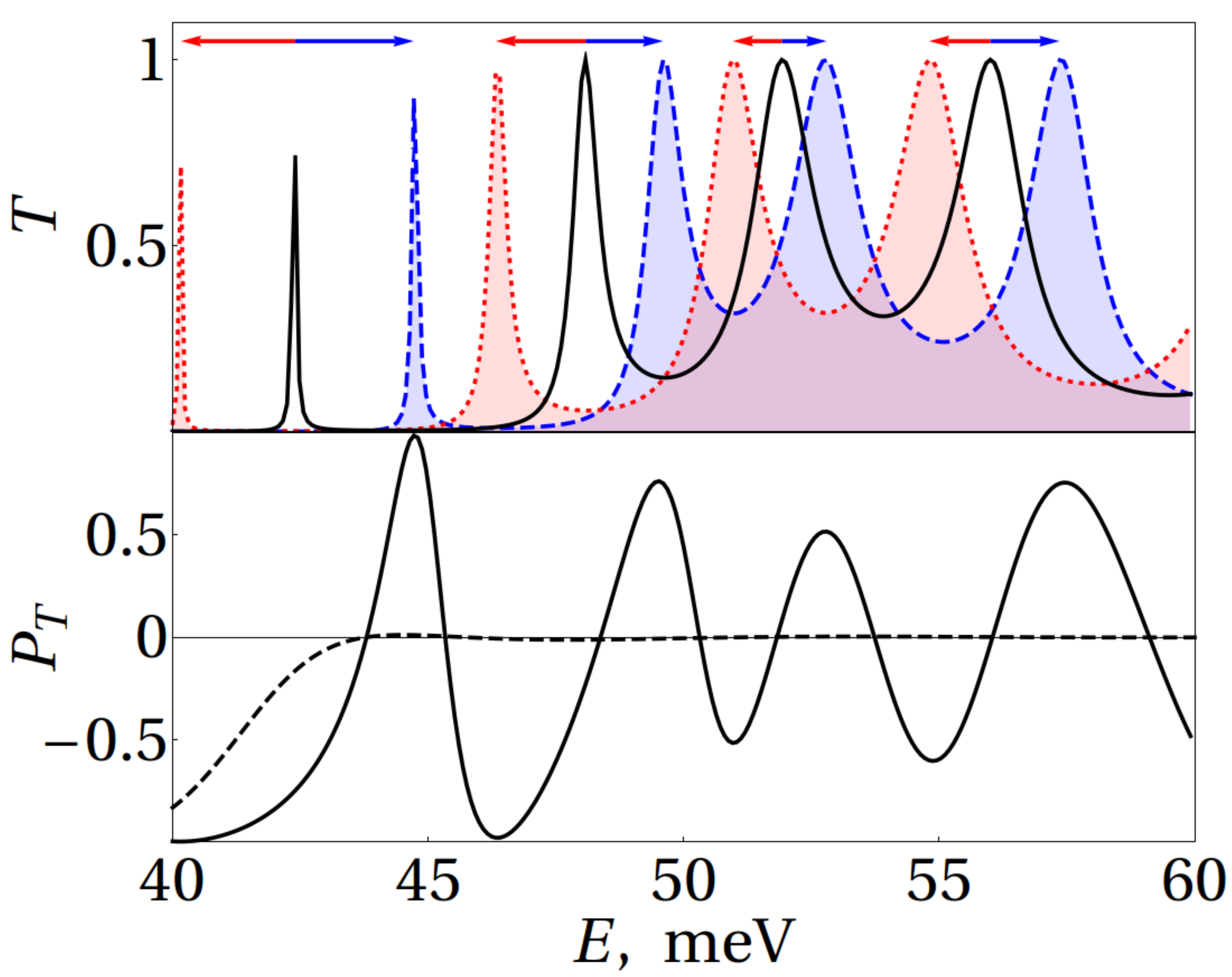}}}
\caption{
Upper panel shows the transmission coefficient calculated without the exchange
splitting (solid black line) and with it (dotted red and dashed
blue lines correspond to $T_{+}$ and $T_{-}$, respectively). Lower panel shows
the degree of the transmission polarization $P_T$ of the nanoring (solid line)
and that of a single nanoribbon (dashed line). All quantities are calculated at
$U_G=0$.
}
\label{figTransmission}
\end{figure}

First, we address the device properties at zero side-gate voltage. 
Transmission coefficients as functions of the carrier energy $E$ are shown
in the upper panel of figure~\ref{figTransmission}.  If the substrate is not
ferromagnetic there is no exchange splitting ($\Delta_\mathrm{ex}=0$) and
the spin does not play any role, so $T_\pm$ are degenerate (see the black
curve in the figure).  In this reference case the transmission is
characterized by a series of peaks or bands which become wider as the
carrier energy increases.  The interaction with the ferromagnet shifts these
features towards lower or upper energies depending on the sign of the
carrier spin (see the red dotted curve giving $T_+$ and the blue dashed one
giving $T_-$).  Note that the transmission spectrum remains qualitatively
the same except for the energy shift.  Peak shifts are shown using arrows
above the curves in the upper panel.

Lower panel of figure~\ref{figTransmission} shows the degree of the
transmission polarization, as defined by equation~(\ref{polarization}),
demonstrating that the transmission is highly polarized within some energy
ranges.  The latter can give rise to the spin polarized electric current. 
The polarization degree is higher at lower energies.  However, the
transmission coefficient in this energy range comprises narrow resonance
peaks which can easily be destroyed by perturbations such as disorder, as we
show below.  The higher energy range with its wider transmission bands,
being more robust under the effects of disorder~\cite{Munarriz11}, is more
promising for applications.  For comparison, we also calculated the
transmission polarization for a single nanoribbon of the same width $w$ and
length $L$ as those of the quantum ring (see the dashed line in the lower
panel of the figure).  As expected, the interference effects are absent and
the transmission polarization disappears quickly as the carrier energy
increases.

\begin{figure}[ht]
\centerline{\resizebox{0.7\columnwidth}{!}{\includegraphics{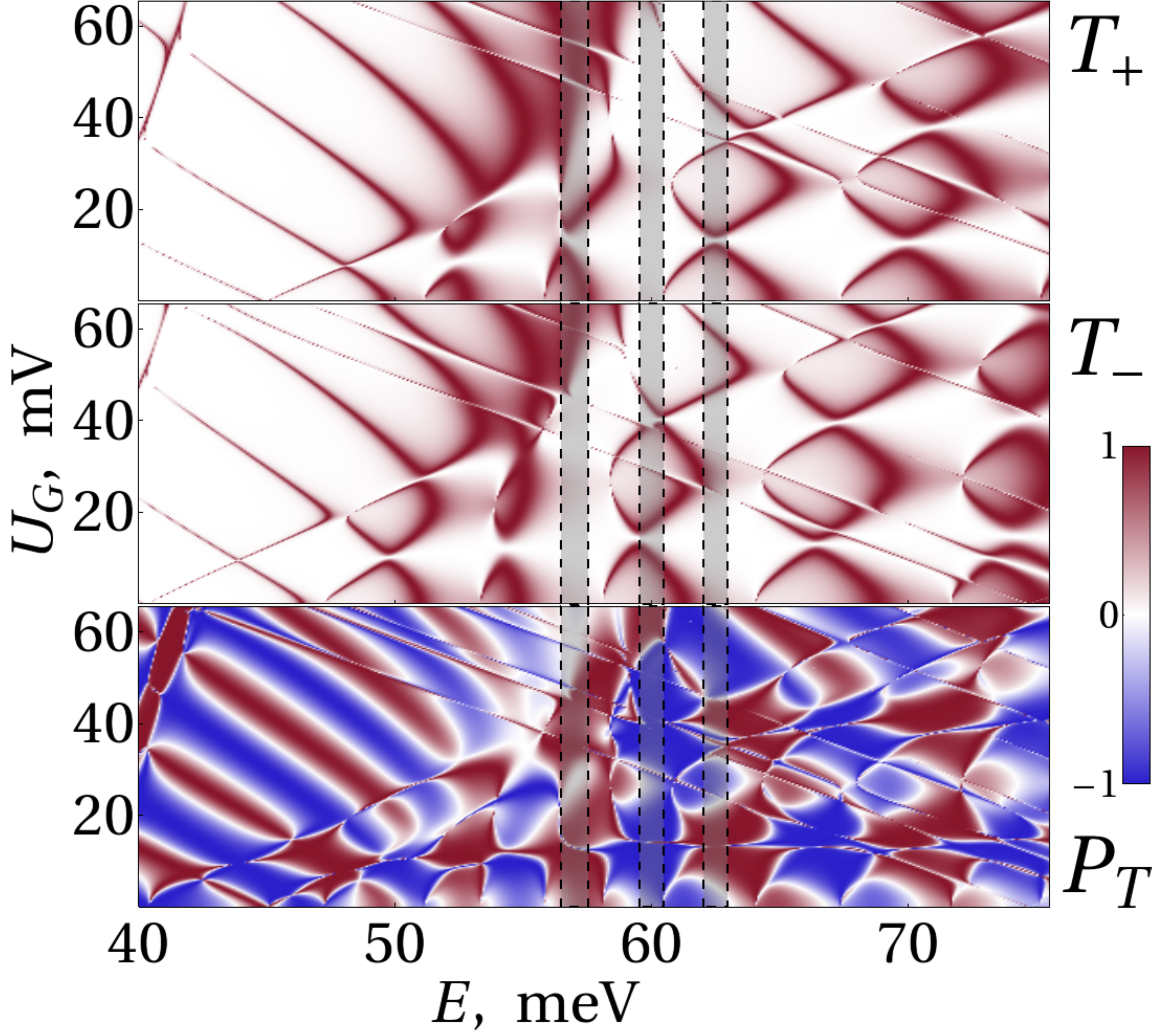}}}
\caption{
Transmission coefficients ${T}_\pm$ and degree of the transmission polarization
$P_T$ as functions of the carrier energy $E$ and the side-gate voltage $U_G$.
}
\label{figPolarizationMap}
\end{figure}

Next, we study the effect of the side gate voltage on the polarization of the electric
current.  To this end, we calculated the transmission coefficient maps for spin
up and spin down electrons. These results are shown in the upper and middle
panels of figure~\ref{figPolarizationMap}. Using them, we obtained the
transmission polarization degree $P_T$ presented in the lower panel of the
figure, which demonstrates that the polarization can be controlled by the
side-gate voltage. Hereafter, we focus on the higher energy range ($E>50$ meV in
the considered case) where wider transmission bands have the interference
nature~\cite{Munarriz11}. As can be seen from the figure, the sign of the
polarization is almost independent of the side-gate voltage for some energies
(see the leftmost and the middle shadowed strips), while for others both the
polarization degree and its sign can be changed by the side-gate (see,
\emph{e.~g.}, the rightmost shadowed strip), which opens the possibility to
control the polarization of the current by the electrostatic gate.

To demonstrate this possibility, we calculated the total current $I$ together
with its polarization degree $P$ according to equations~(\ref{I})
and~(\ref{Ipolarization}).  We assume that Fermi energies of both the source and
the drain are set to some value by a back-gate voltage and then one of these
levels is shifted with respect to the other by the source-drain voltage
$V_\mathrm{SD}$.  Upper and lower limits of the three shadowed strips in
figure~\ref{figPolarizationMap} give the source and drain Fermi energies $\mu_S$
and $\mu_D$, respectively, which were used to calculate electric currents from
equation~(\ref{eqIntensity}) for $V_\mathrm{SD}=1\,$mV and $T=4\,$K.  This
choice of values was suggested by the following reasoning.  At $T=0$ the Fermi
distributions are step functions, so only the energy range between $\mu_S$ and
$\mu_D$ contribute to the current [see the integrand in
equation~(\ref{eqIntensity})]. If this range is greater than the typical energy
separation between the transmission bands of $T_{+}$ and $T_{-}$ (see
figure~\ref{figPolarizationMap}), then both polarizations would contribute to
the total current to a similar extent and its polarization would be reduced. 
Therefore, to observe substantial current polarization, $e V_\mathrm{SD}$
should be smaller than the typical energy scale $\Delta E$ in the transmission
maps.  Figure~\ref{figPolarizationMap} suggests that this scale is of the order
of $5\,$meV, which justifies our choice of $V_\mathrm{SD}=1\,$mV. Similar
arguments apply to the temperature which smears out the Fermi step functions,
increasing the range of energies contributing to the current and reducing
its polarization.  Thus, the
temperature should be smaller than $\Delta E/k_B \approx 60\,$K.

\begin{figure}[ht]
\centerline{\resizebox{0.5\columnwidth}{!}{\includegraphics{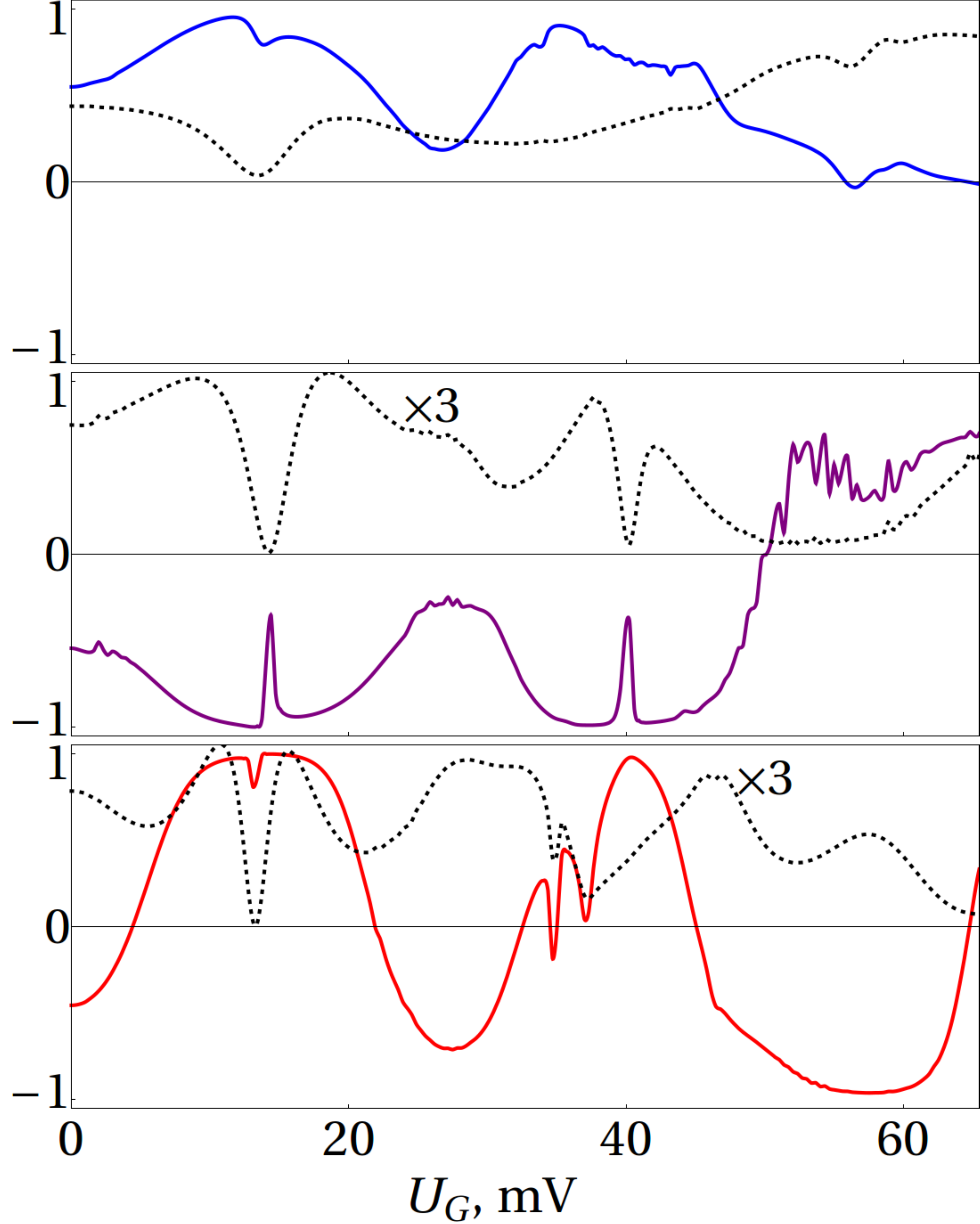}}}
\caption{
Total electric current $I$ through the device (dotted lines) and its
polarization degree (solid lines) as functions of the side-gate voltage $U_G$.
$I$ is given in units of the current obtained for the case of a perfectly
transmitting sample [$T_{\pm}(E,U_G)\equiv1$]. Upper, middle and lower panels
correspond  to the leftmost, middle and rightmost shadowed strips in 
figure~\ref{figPolarizationMap} (see text for details).
}
\label{figIntensities}
\end{figure}

Figure~\ref{figIntensities} shows the total current $I$
(dotted lines) and its polarization degree $P$ (solid lines). Upper, middle and
lower panels correspond to the leftmost, middle and rightmost shadowed strips in
figure~\ref{figPolarizationMap}, respectively. As expected, the electric current
can be highly polarized. For some combinations of the source and drain Fermi
energies, the sign of the current polarization remains the same within wide
ranges of the side-gate voltage (see the upper and the middle panels of
figure~\ref{figIntensities}). Nevertheless, as can be seen from the lower panel,
the Fermi energies at the source and drain can be adjusted in such a way that
the current polarization can be changed in almost its entire possible range
$[-1,1]$ by the side-gate voltage, suggesting that the proposed device operates
as a controllable source of spin-polarized electrons.

We note that we have considered an ideal symmetric nanoring while different
imperfections or perturbations, in particular, the disorder, can affect the electric
current and its polarization.  There are various possible sources of disorder,
for example, charged impurities in the substrate or defects of the device
fabrication, such as imperfections of the device edges. While the former would
provide some additional smooth electrostatic potential and can hardly
deteriorate the transmission through the device to a large extent, the impact of
the latter on the transport properties is probably stronger, especially for
small devices. 

\begin{figure}[ht]
\centerline{\resizebox{0.6\columnwidth}{!}{\includegraphics{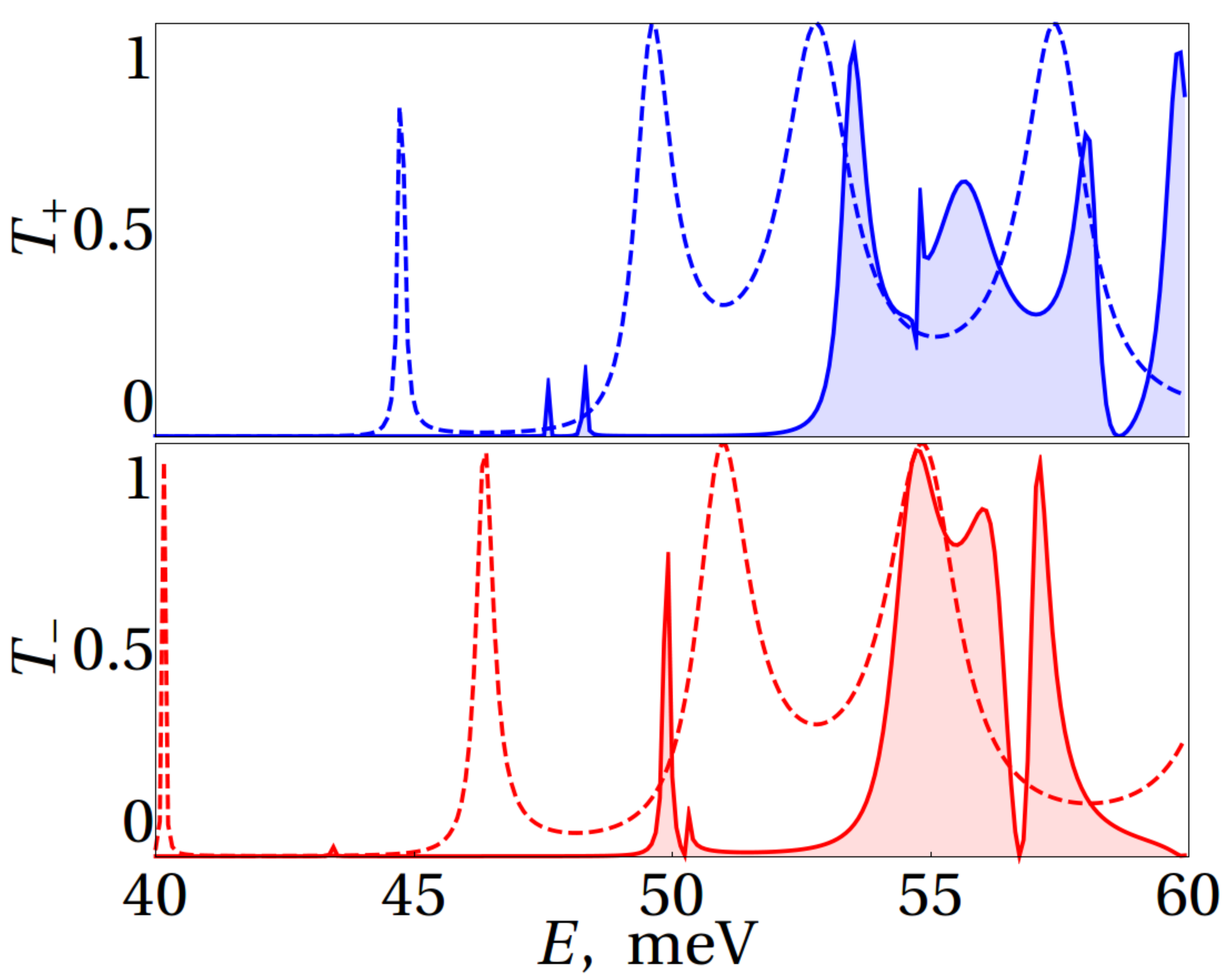}}}
\caption{
Transmission coefficients $T_{\pm}$ of the regular sample (dashed lines) and
those of a disordered one (solid curves with filling). The latter are
calculated for one realization of the edge disorder. The side-gate voltage
$U_G$ is set to zero.
}
\label{figDisorder}
\end{figure}

Following Ref.~\cite{Munarriz11}, in order to estimate a possible
impact of the edge disorder on the transport properties, we consider a sample in
which we remove pairs of carbon atoms from the nanoring edges with some given
probability $p$. 
Transmission coefficients $T_{\pm}$ calculated for one particular realization of
such a disorder for $p=0.05$ and zero side-gate voltage are presented in the
upper and lower panels of figure~\ref{figDisorder}, respectively.  Dashed lines
show the transmission coefficients of the reference ordered sample while
solid
lines show those of a disordered one. In the disordered sample all transmission
bands are shifted to higher energies with respect to their positions in the
regular one because, by removing atoms from the edges, the ribbons are made
effectively narrower. This leads to higher quantization energy in the lateral
direction that manifests itself in the observed shifts. As expected, narrow
resonance peaks in the lower energy range ($E<50\,$meV) almost disappear in the
disordered sample while wider interference-related bands at higher energies are
not destroyed by the disorder.  These bands are affected by the disorder
to a comparable degree for both spin up and spin down electrons, which suggests
that a moderate disorder would not deteriorate polarization properties of the
spin filter to a large extent.

Finally, we address possible effects of a different kind of imperfections of the
device fabrication on its performance. One of them is a change of the length of
one of the arms at a fixed nanoribbon width $w$.  Such a change would affect the
interference at the drain, and one can expect the transmission peaks
appearing at different side-gate voltages, while the overall picture remaining
qualitatively the same. We have checked the validity of this intuitive
argumentation by numerical calculations (results are not presented here).

\begin{figure}[ht]
\centerline{\resizebox{0.6\columnwidth}{!}{\includegraphics{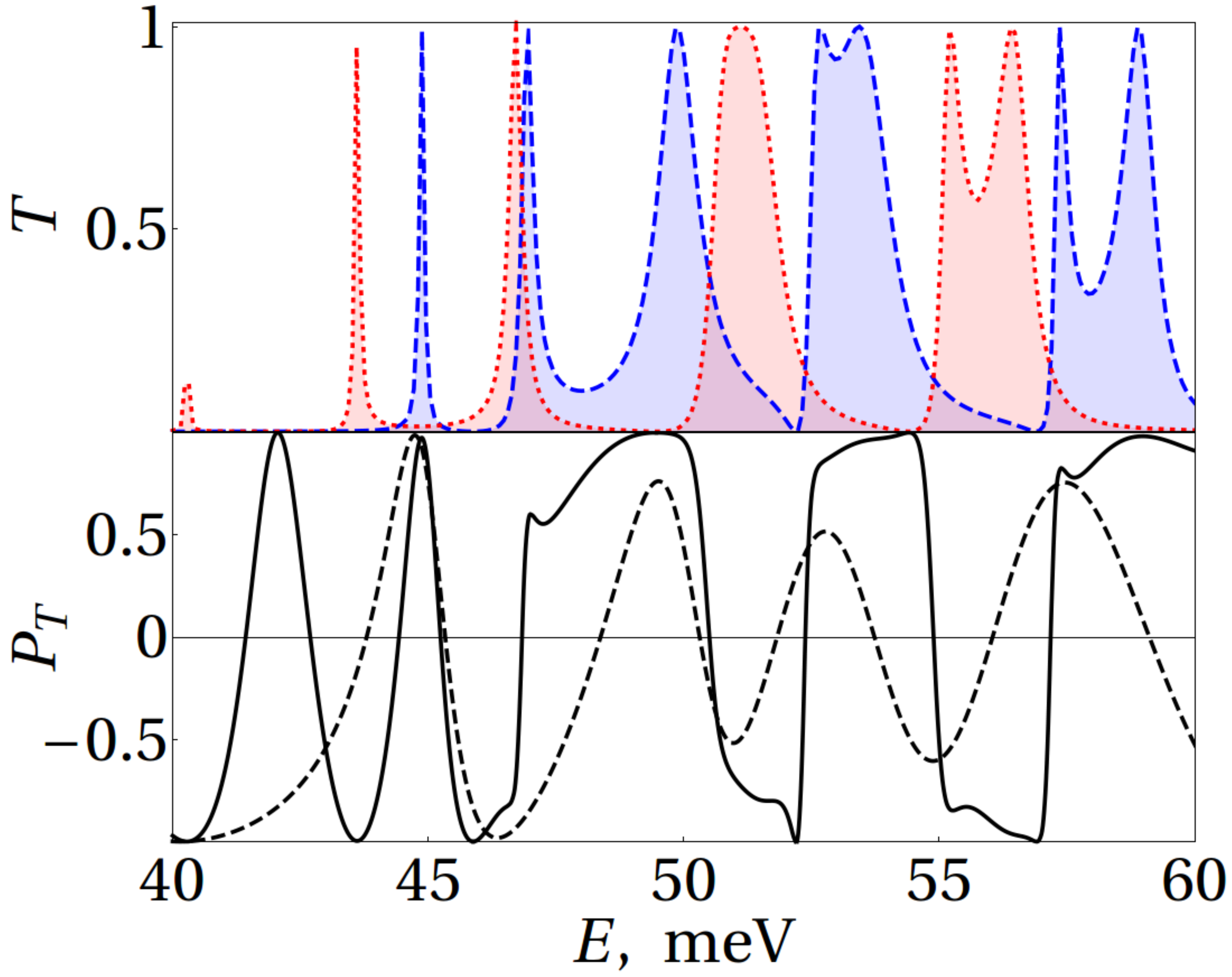}}}
\caption{
Upper panel shows the transmission coefficient for the asymmetric nanoring
with an extra layer of atoms in the horizontal part of the upper arm (dotted
red and dashed blue lines shows $T_{+}$ and $T_{-}$, respectively).  Lower
panel displays the transmission polarization for the symmetric nanoring
(dashed line) and the asymmetric one (solid line).  All quantities
are calculated for $U_G=0$.
}
\label{figAsymRing}
\end{figure}

The fabrication imperfection that could have a more profound impact on
transport is a change of the width of one of the arms.  In this case, the
lateral quantization in this wider (or narrower) part of the ring is
different, resulting in a mode mismatch at the boundary, which can reduce
the transmission.  In order to estimate the corresponding effect, we added
an extra layer of atoms to the horizontal part of the upper arm and
calculated $T_{\pm}$ and $P_T$.  The results are presented in
figure~\ref{figAsymRing}, where the upper panel shows the transmissions
$T_{\pm}$ for the asymmetric sample at zero side-gate voltage, $U_G=0$. 
Comparing the transmission coefficients in the upper panel of the figure
with those in figure~\ref{figTransmission} one can see that, as can be
expected, the asymmetry shifts the transmission peaks.  Besides, the
overlaps between the transmission bands are reduced in the case of the
asymmetric nanoring.  The bands are more isolated, which results in the most
important effect: the polarization sign can be switched much more abruptly,
making the asymmetric design more advantageous for applications.

To conclude the discussion of the impact of fabrication imperfections on the
device properties, we note that, because the interference is very sensitive to
many details, different samples could be expected to have quantitatively
different current-voltage characteristics.  However, as we have shown above,
qualitatively the transmission coefficient and its polarization remains the
same. It can then be expected that as long as the size of the ring, temperature,
and applied voltages are such that the system remains within the one-mode
regime, the device can operate as a tunable spin filter.

\section{Summary}

In summary, we have proposed and studied a novel spin filter which exploits
quantum interference effects.  The device comprises a graphene nanoring with
60$^\circ$ bends fabricated above a ferromagnetic strip.  We showed that due to
the exchange splitting induced by the magnetic ion of the ferromagnetic layer,
the transmission coefficient is different for spin up and spin down electrons,
giving rise to the polarization of the conductance and the electric current.  We
demonstrated that both the current and its polarization can be controlled by a
side-gate voltage.  Predicted effects are shown to be robust under a moderate
edge disorder and other fabrication imperfections, such as the asymmetry of the
ring.  Therefore, we conclude that the proposed device is a promising candidate
for real world applications, in particular, it can be used as a tunable source
of polarized electrons.


Work at Madrid was supported by MICINN (Projects Mosaico and MAT2010-17180). 
P.\ A.\ O.\ acknowledges support from FODNECYT (Project 1100560).

\end{document}